\documentclass[pdftex,12pt]{JHEP3}
\usepackage{amsmath,amsfonts,amsbsy,latexsym,amssymb,amscd,amstext}
\usepackage[pdftex]{epsfig}
\pdfoutput=1
 \def\be{\begin{equation}}
\def\ee{\end{equation}}
\def\bma{\begin{pmatrix}}
\def\ema{\end{pmatrix}}
\def\bea{\begin{eqnarray}}
\def\eea{\end{eqnarray}}
\def\pd{\partial}
\def\a{\alpha}
\def\b{\beta}
\def\g{\gamma}

\def\m{\mu}
\def\n{\nu}

\def\l{\lambda}

\def\bi{\begin{itemize}}
\def\ei{\end{itemize}}

\date{May 25th, 2008} \preprint{FTUAM-10-25; IFT-UAM/CSIC-10-56}
\title{Bottom up approach to Quantum Gravity } \author{Enrique \'Alvarez  \\  Instituto de F\'{\i}sica Te\'orica
UAM/CSIC and Departamento de F\'{\i}sica Te\'orica \\ Universidad
Aut\'onoma de Madrid, E-28049--Madrid, Spain \\ E-mail: \email{enrique.alvarez@uam.es }}
\abstract{
A general introduction is given to what  can be predicated about quantum gravity once the lessons from the standard model of particle physics are taken into account. In particular, the effective lagrangian point of view is briefly commented upon .

}
\begin{document}

{\vskip 1cm}
\newpage
\section {Introduction}
Can we assert some  proposition about quantum gravity with some confidence? 
\par

\par
The mass scale associated to this problem just by sheer dimensional analysis is
Planck's mass, which is given in terms of Newton's constant, $G$ by
\be
m_p\equiv\sqrt{\hbar c\over 8\pi G}\sim G^{-1/2}\sim 10^{19}GeV
\ee 
(we shall always work in units so that $\hbar=c=1$).   
If we remember that $1\,GeV(=10^3\, MeV)$ is the rough scale of hadronic physics (the mass and inverse Comptom wavelength of a proton, for example), this means that quantum gravity effects will only be apparent when we are able to explore concentrated energy  roughly $10^{19}$ times bigger (or an scale distance correspondingly smaller; these two statements are supposed to be equivalent owing to Heisenberg's principle). To set the scale, the Large Hadron Collider works roughly at the $TeV(=10^3\,GeV)$ scale, so there is a long way to go before reaching expected quantum gravity effects in accelerators.
\par
If we want to get experimental information, we have to turn our attention towards Cosmology, or perhaps look for some clever precison experiment in the laboratory. 
Lacking any experimental clue, the only thing we can do is to think and try to look for logical (in)consistencies.  
\par
It has been repeatedly argued by many particle physicists that the practical utility of the answer to this question will not presumably be great. How would we know for sure beforehand?. There has always been a recurrent dream, exposed vehemently by Salam \cite{Salam} that the inclusion of the gravitational interaction would cure many of the diseases and divergences of quantum field theory, through the inclusion in the propagator of terms of the type
\[
e^{-{1\over m_p^2 x^2}}
\]
So that for example, the sum of tree graphs that leads to the Schwarzschild solution as worked out by Duff \cite{Duff}
\[
{1\over r}+ {2M\over m_p^2 r^2}+{4 M^2\over m_p^4 r^3}+\ldots
\]
would get modified to
\[
{1\over r}e^{-{1\over m_p r}}+ {2M\over m_p^2 r^2}e^{-{ 2 \over m_p r}}+\ldots\sim {1\over r e^{-{1\over m_p r}}-2	{M\over m_p^2}}
\]
shifting the location of the horizon and eliminating the singularity at $r=0$. Nobody has been able to substantiate this dream so far.
\par

 Ay ant rate quantum gravity is nevertheless a topic which has fascinated whole generations of physicists, just because it is so difficult.
There seems to be a strong tension between the beautiful, geometrical world of General Relativity and the no less marvelous, less geometrical, somewhat mysterious, but very well tested experimentally, world of Quantum Mechanics.

As with all matters of principle  we can hope to better understand both quantum mechanics and gravitation if we are able to clarify the issue.
\par
The most conservative approach is of course to start from what is already known with great precision about the standard model of elementary particles associated to the names of Glashow, Weinberg and Salam. This can be called the {\em bottom-up approach} to the problem.
In this way of thinking Wilson taught us that there is a working low energy effective theory, and some quantum effects in gravity can be reliably computed for energies much smaller than Planck mass. There are two caveats to this. First of all, we do not understand why the observed cosmological constant is so small: the natural value from the low energy effective lagrangian point of view ought to be much bigger.  The second point is that one has to rethink again the lore of effective theories in the presence of horizons. 
We shall comment on both issues in due time.
\par
There is not a universal consensus even on the most promising avenues of research from the opposite {\em top-down} viewpoint. Many people think that strings \cite{Strings} are the best
buy (I sort of agree with this); but it is true that after more than two decades of intense effort nothing substantial has come out of them. Others \cite{Rovelli} try to quantize directly the Einstein-Hilbert lagrangian, something that is at variance with our experience in effective field theories. But it is also true that as we have already remarked, the smallish value of the observed cosmological constant also cries out of the standard effective theories lore.
\section{Symmetries and observables}
It is generally accepted that General Relativity, a generally covariant theory, is akin to a gauge theory, in the sense that the diffeomorphism group of the apace-time manifold, $Diff(M)$ plays a role similar to the compact gauge group in the standard model of particle physics. There are some differences though. To begin with, the group, $ Diff(M)$ is too large; is not even a Lie group \cite{Milnor}. Besides, its detailed structure depends on the manifold, which is a dynamical object not given a priori. Other distinguished subgroups (such as the area-preserving diffeomorphisms,\cite{Alvarezz}) are perhaps also arguable for. Those leave invariant a given measure, such as the Lebesgue measure, $d^n x$.
\par
It also seems clear that when there is a boundary of space-time, then the gauge group is restricted to the subgroup consisting on those diffeomorphisms that act trivially on the boundary. The subgroup that act not-trivially is related to the set of conserved charges. In the asympotically flat case this is precisely the Poincar\'e group, $SO(1,3)$ that gives rise to the ADM mass. 
\par
In the asympotically anti-de Sitter case, this is presumably related to the conformal group $SO(2,3)$. 
\par
It is nevertheless not clear what is the physical meaning of keeping constant the boundary of spacetime (or keeping constant some set of boundary conditions) in a functional integral of some sort.
\par
A related issue is that
 it is very difficult to define what could be {\em observables} in a diffeomorphism invariant theory, other than global ones defined as integrals of scalar composite operators $O(\phi_a(x))$ (where $\phi_a, a=1\ldots N$ parametrizes all physical fields) with the peudo-riemannian measure
\[
{\cal O}\equiv \int \sqrt{|g|}d^4 x O(\phi_a(x))
\]
 Some people claim that there are no local observables whatsoever, but only {\em pseudolocal } ones \cite{Giddings}; the fact is that we do not know.
 Again, the exception to this stems from keeping the boundary conditions fixed; in this case it is possible to define an $S$-matrix in the asymptotically flat case, and a conformal quantum field theory (CFT) in the asymptotically anti-de Sitter case. 
 Unfortunatelly, the most interesting case from the cosmological point of view, which is when the space-time is asymptotically de Sitter is not well understood.
\par
Incidentally, it is well known that the equivalence problem in four-dimensional geometries is undecidable \cite{Alvarezzz}. In three dimensions  Thurston's geometrization conjecture has recently been put on a firmer basis by Hamilton and Perelman, but it is still  not clear whether it can be somehow implemented in a functional integral without some drastic restrictions. Those caveats should be kept in mind when reading the sequel. 
\par
A radically different viewpoint has recently been advocated by 't Hooft \cite{'tHooft} by insisting in causality to be well-defined, so that the conformal class of the space-time metric should be determined by the physics, but not necessarily the precise point in a given conformal orbit. If we write the spacetime metric in terms of a unimodular metric and a conformal factor
\[
g_{\m\n}=\omega^2(x)\hat{g}_{\m\n}
\]
with
\[
det\,\hat{g}_{\m\n}=1
\]
then the unimodular metric is in some sense intrisic and determines causality, whereas the conformal factor depends on the observer in a way dictated by {\em black hole complementarity}.
\par
Finally, there is always the (in a sense, opposite) possibility that space-time (and thus diffeomorphism invariance) is not a fundamental physical entity in such a way that the appropiate variables for studying short distances are non geometrical. Something like that could happen in string theory, but our understanding of it is still in its infancy.

\section{The Effective Lagrangian Approach to Quantum Gravity}

The main purpose of this talk is to review the least ambitious approach possible to the topic. Actually, one can discuss what is the notion of causality when the spacetime metric is a fluctuating object, or whether there is an adequate notion of time to be used in writing down Scr\"odinger's equation for the Universe itself. But if our previous experience with the other interactions is to be of any relevance here, there ought to be a regime, experimentally accessible in the not too distant future, in which gravitons propagating in flat spacetime can be isolated. This is more or less unavoidable, provided gravitational waves are discovered experimentally, and the road towards gravitons should not be too different from the road that lead from the discovery of electromagnetic waves to the identifications of photons as the quanta of the corresponding interaction, a road that led from Hertz to Planck. 
\par
Any quantum gravity theory that avoids identifying gravitational radiation as consisting of large numbers of gravitons in a semiclassical state would be at variance with all we believe to know about quantum mechanics. 
\par
What we expect instead to be confirmed by observations somewhere in the future is that the number of gravitons per unit volume with frequencies between 
$\omega$ and $\omega+d\omega$ is given by Planck's formula
\[
n(\omega)d\omega={\omega^2 \over \pi^2}{1\over e^{\hbar \omega\over k T}-1} d\omega
\]

It is natural to keep an open mind for surprises here, because it can be argued that gravitational interaction is not alike any other fundamental interaction in the sense that the whole structure of space-time ought presumably be affected, but it cannot be denied that this is the most conservative approach and as such it should be explored first, up to its very limits, which could hopefully indicate further avenues of research.

From our experience then with the standard model of elementary particles, and assuming we have full knowledge of the fundamental symmetries of our problem, we know that we can parametrize our ignorance on the
{\em fundamental} ultraviolet (UV) physics by writing down all local operators in the low energy fields $\phi_i(x)$ compatible with the basic symmetries we have assumed.
\[
L=\sum_{n=0}^\infty{\l_n(\Lambda)^n\over \Lambda^n}{\cal O}^{(n+4)}\left(\phi_i\right)
\]
Here $\Lambda$ is an ultraviolet (UV) cutoff, which restricts the contributions of large euclidean momenta (or small euclidean distances) and $\l_n(\Lambda)$ is an infinite set of dimensionless bare couplings.
\par
Standard Wilsonian arguments imply that {\em irrelevant operators}, corresponding to $n > 4$, are less and less important as we are interested in deeper and deeper infrared (IR) ({\em low energy}) variables.  The opposite occurs wuth {\em relevant operators}, corresponding to $n<4$, like the masses, that become more and more important as we approach the IR. 
The intermediate role is played by the {\em marginal operators}, corresponding to precisely $n=4$, and whose relevance in the I R is not determined solely by dimensional analysis, but rather by quantum corrections. 
The range of validity of any finite number of terms in the expansion is roughly
\[
{E\over\Lambda}<< 1
\]
where $E$ is a characteristic energy of the process under consideration.
\par
In the case of gravitation, we assume that general covariance (or diffeomorphism invariance) is the basic symmetry characterizing the interaction. We can then write

\bea
&&L_{eff}=\l_0 \Lambda^4 \sqrt{|g|}+\l_1 \Lambda^2 R \sqrt{|g|}+\l_2 R^2+ {1\over 2}g^{\a\b}\nabla_\a\phi\nabla_\b\phi\sqrt{|g|}+\nonumber\\
&&+\l_3 {1\over \Lambda^2}R^{\a\b}\nabla_\a\phi\nabla_\b\phi\sqrt{|g|}+\l_4 {1\over \Lambda^2}R^3\sqrt{|g|}+\l_5 \phi^4\sqrt{|g|}+\nonumber\\
&&+\bar{\psi}\left(e^\m_a \g^a \left(\pd_\m-\omega_\m\right)\psi-m\right)\psi+{\l_5\over \Lambda^2}\bar{\psi}e^\m_a \g^a R\left(\pd_\m-\omega_\m\right)\psi+\ldots
\eea
where $e_a^\m$ is the tetrad, such that
\[
e_a^\m e_\b^\n \eta^{\a\b}=g^{\m\n}
\]
$\eta^{\a\b}$ being Minkowski's metric. The quantities $\omega_\m$ are the spin connection.
\par
The need to recover General Relativity in the classical IR limit means 
\[
\l_1\Lambda^2=-{c^3\over 16\pi G}\equiv -2 M_p^2
\]
This in turn, means that if
\[
\l_0\Lambda^4
\]
is to yield the observed value for the cosmological constant (which is of the order of magnitude of Hubble's constant, $H_0^4$, which is a very tiny figure when  expressed in particle  physics units, $H_0\sim 10^{-33}\,eV$) then
\[
\l_0\sim 10^{-244}
\]
This is one aspect of the cosmological constant problem; it seems most unnatural that the cosmological constant is observationally so small from the effective lagrangian point of view. I do not have anything new to say on this.
\par
This expansion is fine as long as it is considered a low energy expansion. As Donoghue \cite{Donoghue} has emphasized, even if it is true that each time that a renormalization is made there is a finite arbitrariness, there are physical predictions stemming from the non-local finite parts.
\par
The problem is when energies are reached that are comparable to Planck's mass,
\[
E\sim M_p.
\]
Then all couplings  in the effective Lagrangian become of order unity, and there is no {\em decoupling limit} in which gravitation can be considered by itself in isolation from all other interactions.
This then seems the minimum prize one has to pay for being interested in quantum gravity; all couplings in the derivative expansion become important simultaneously. No significant differences appear when supergravity is considered.
\par 
In conclusion, it does not seem likely that much progress can be made by somehow quantizing Einstein-Hilbert's Lagrangian in isolation. To study quantum gravity means to study all other interactions as well.
\par
On the other hand, are there any reasons to go beyond the standard model (SM)? 
\par
Yes there are some, both theoretical, and experimental. From the latter, and most important, side, both the existence of neutrino masses and  dark matter do not fit into the SM. And from the former, abelian sectors suffer from Landau poles and are not believed to be UV complete; likewise the self-interactions in the Higgs sector appear to be a trivial theory. Also the experimental values of the particle masses in the SM are not natural from the effective lagrangian point of view.
\par
The particle physics community has looked thoroughly for such extensions since the eighties: extra dimensions (Kaluza-Klein), supersymmetry and supergravity, technicolor, etc. From a given point of view, the natural culmination of this road is string theory 
\par
A related issue is the understanding of the so-called {\em semiclassical gravity}, in which the second memnber of Einstein's equations is taken as the expectation value of some quantum energy-momentum operator. It can be proved that this is the dominant $1/N$ approximation in case there are $N$ identical matter fields (confer \cite{Hartle}). In spite of tremendous effort, there is not yet a full understanding of Hawking's emission of a black hole from the effective theory point of view (confer, for example, \cite{Hooft}). Another topic in which this approach has been extensively studied is Cosmology. Novel effects (or rather old ones on which no emphasis was put until recently) came from lack of momentum conservation and sem to point towards some sort of instability \cite{Polyakov}; again the low energy theory is not fully understood; this could perhaps have something to do with the presence of horizons.

\par
Coming back to our theme, and  closing the loop, what are the prospects to make progress in quantum gravity? 
\par

Insofar as effective lagrangians are a good guide to the physics there are only two doors open: either there is a ultraviolet (UV) attractive fixed point in coupling space, such as in Weinberg\'s {\em asymptotic safety} \cite{Weinberg} or else new degrees of freedom, like in string theory \cite{Strings} exist in the UV.
Even if Weinberg\'s approach is vindicated, the fact that the fixed point most likely lies at strong coupling combined with our present inability to perform analitically other than perturbative computations, mean that lattice simulations should be able to cope with the integration over (a subclass of) geometries before physical predictions could be made with the techniques at hand at the present moment.
\par

 It is to be remarked that sometimes theories harbor the seeds of their own destruction. Strings for example, begin as theories in flat spacetime, but there are indications that space itself should be a derived, not fundamental concept. It is hoped that a simpler formulation of string theory exists bypassing the roundabouts of its historical development. This is far from being the case at present.

\par

 Finally, it is perhaps worth pointing out that to the extent that a purely gravitational canonical approach, as the ones based upon the use of Ashtekar \cite{Ashtekar} variables makes contact with the classical limit (which is an open problem from this point of view)  the preceding line of argument should  still  carry on. 
 \par
 It seems {\em unavoidable} with our present understanding, that any theory of quantum gravity should recover, for example, the prediction that there are quantum corrections to the gravitational potential given by
 \cite{BjerrumBohr}
 \[
 V(r)=-{G m_1 m_2 \over r}\left(1+3 {G\left(m_1+m_2\right)\over r}+{41 \over 10\pi}{G\hbar\over r^2}\right)
 \]

(the second term is also a loop effect, in spite of the conspicuous absence of $\hbar$.)
Similarly, and although this has been the subject of some controversy, it seems now established that there are gravitational corrections to the running of gauge couplings, first uncovered by Robinson and Wilczek \cite{Robinson} and given in standard notation by
\[
\b(g,E)=-{b_0\over (4\pi)^2}g^3 -3 {16\pi G\over (4\pi)^2 \hbar c^3}g E^2
\]
Sometimes these effects are dismissed as perturbative, and therefore trivial. This is not a healthy attitude.


\section*{Acknowledgments}
I am grateful to Jaume Garriga, Tini Veltman and Enric Verdaguer for many discussions, as well as to  John Donoghue, David Toms, and Sigurdur Helgason for useful correspondence. I also would like to thank the organizers of ERE 2010, C. Barcel\'o and J.L. Jaramillo, for the kind invitation to attend this sucessful meeting.

This work has been partially supported by the
European Commission (HPRN-CT-200-00148) as well as by FPA2009-09017 (DGI del MCyT, Spain) and 
 S2009ESP-1473 (CA Madrid).

\newpage



\begin{thebibliography}{99}

\bibitem{Alvarez}
E.~Alvarez,
``Quantum Gravity: A Pedagogical Introduction To Some Recent Results,''
Rev.\ Mod.\ Phys.\  {\bf 61} (1989) 561.

\bibitem{Alvarezz}
E. Alvarez, ``Can one tell Einstein's unimodular theory from Einstein's general
  JHEP {\bf 0503} (2005) 002
  [arXiv:hep-th/0501146].\\

\bibitem{Alvarezzz}
  E.~Alvarez,
  ``Some general problems in quantum gravity,''\\
``Some general problems in quantum gravity. 2. The Three-dimensional case,''
  Int.\ J.\ Mod.\ Phys.\  D {\bf 2} (1993) 1
  [arXiv:hep-th/9211050].

\bibitem{BjerrumBohr}
  N.~E.~JBjerrum-Bohr, J.~F.~Donoghue, B.~R.~Holstein,
  ``Quantum gravitational corrections to the nonrelativistic scattering potential of two masses,''
  Phys.\ Rev.\  {\bf D67}, 084033 (2003).
  [hep-th/0211072].



\bibitem{Ashtekar}
  A.~Ashtekar,
  ``New Variables for Classical and Quantum Gravity,''
  Phys.\ Rev.\ Lett.\  {\bf 57}, 2244 (1986).

\bibitem{dewitt} B. S. de Witt,{\em Dynamical Theory of Groups and Fields},
                  (Gordon and Breach, 1965)\\{\em Quantum theory of gravity, I,II,III}
                   Phys. Rev 162 (1967)1113,1195,1239.
                   
                   
                   
                   
\bibitem{Donoghue} John F. Donoghue, {\em General Relativity as an effective field 
theory: The leading quantum
                   corrections}, Phys. Rev. D50 (1994) 3874.\\
  ``Perturbative dynamics of quantum general relativity,''
  arXiv:gr-qc/9712070.\\
%

\bibitem{Duff}
  M.~J.~Duff,
  ``Quantum Tree Graphs and the Schwarzschild Solution,''
  Phys.\ Rev.\  D {\bf 7} (1973) 2317.


\bibitem{Feynman}R.P. Feynman, {\em Lectures on Gravitation},
                 (University of Bangalore Press,1997)

\bibitem{Giddings}
  S.~B.~Giddings, D.~Marolf and J.~B.~Hartle,
  ``Observables in effective gravity,''
  Phys.\ Rev.\  D {\bf 74} (2006) 064018
  [arXiv:hep-th/0512200].

\bibitem{Hartle}
  J.~B.~Hartle, G.~T.~Horowitz,
  ``Ground State Expectation Value Of The Metric In The 1/n Or Semiclassical Approximation To Quantum Gravity,''
  Phys.\ Rev.\  {\bf D24}, 257-274 (1981).
  
\bibitem{Hooft}
  G.~'t Hooft,
  ``The Scattering matrix approach for the quantum black hole: An Overview,''
  Int.\ J.\ Mod.\ Phys.\  {\bf A11}, 4623-4688 (1996).
  [gr-qc/9607022].

\bibitem{Milnor}
  J.~Milnor,
  ``Remarks On Infinite Dimensional Lie Groups,''
{\it  In *Les Houches 1983, Proceedings, Relativity, Groups and Topology, Ii*, 1007-1057}

\bibitem{Polyakov}
 A.~M.~Polyakov,
 ``De Sitter Space and Eternity,''
 Nucl.\ Phys.\  B {\bf 797} (2008) 199
 [arXiv:0709.2899 [hep-th]].\\
``Decay of Vacuum Energy,''
  Nucl.\ Phys.\  B {\bf 834}, 316 (2010)
  [arXiv:0912.5503 [hep-th]].
  E.~Alvarez, R.~Vidal,
  ``Comments on the vacuum energy decay,''
   [arXiv:1004.4867 [hep-th]].


\bibitem{Robinson}
  S.~P.~Robinson, F.~Wilczek,
  ``Gravitational correction to running of gauge couplings,''
  Phys.\ Rev.\ Lett.\  {\bf 96}, 231601 (2006).
  [hep-th/0509050].\\
  D.~J.~Toms,
  ``Quantum gravitational contributions to quantum electrodynamics,''
   [arXiv:1010.0793 [hep-th]].



\bibitem{Rovelli}
  C.~Rovelli,
  ``Quantum Gravity,''
{\it  Cambridge, UK: Univ. Pr. (2004) 455 p}


\bibitem{Strings}
Two standard references are\\
 M.~B.~Green, J.~H.~Schwarz and E.~Witten,
  ``SUPERSTRING THEORY. VOL. 1: INTRODUCTION,''
{\it  Cambridge, Uk: Univ. Pr. ( 1987) 469 P. ( Cambridge Monographs On Mathematical Physics)}\\
  ``Superstring Theory. Vol. 2: Loop Amplitudes, Anomalies And Phenomenology,''
{\it  Cambridge, Uk: Univ. Pr. ( 1987) 596 P. ( Cambridge Monographs On Mathematical Physics)}\\
J.~Polchinski,
  ``String theory. Vol. 1: An introduction to the bosonic string,''
{\it  Cambridge, UK: Univ. Pr. (1998) 402 p}\\
 ``String theory. Vol. 2: Superstring theory and beyond,''
{\it  Cambridge, UK: Univ. Pr. (1998) 531 p}\\
A shorter reference is:
  E.~Alvarez and P.~Meessen,
  ``String primer,''
  JHEP {\bf 9902}, 015 (1999)
  [arXiv:hep-th/9810240].


\bibitem{Salam}
  A.~Salam,
  ``Impact Of Quantum Gravity Theory On Particle Physics,''




\bibitem{'tHooft}
  G.~'t Hooft,
  ``Quantum gravity without space-time singularities or horizons,''
  arXiv:0909.3426 [gr-qc].



\bibitem{Weinberg} S. Weinberg,{\em Ultraviolet divergences in quantum theories of gravitation},
                     in {\em General relativity: an Einstein centenary volume}, 
                     S.W. Hawking and W. Israel eds.
                     (Cambridge University press,1979).\\
                     ``Effective Field Theory, Past and Future,''
  arXiv:0908.1964 [hep-th].








\end{thebibliography}
\end{document}